\documentclass[prl,aps,twocolumn,showpacs,superscriptaddress,notitlepage]{revtex4-1}

\usepackage{graphicx}
\usepackage{amsmath}
\usepackage{amssymb}
\usepackage{times}
\usepackage{color}
\usepackage{bbm}
\usepackage{bm}
\usepackage{xcolor}
\usepackage{soul}
\usepackage[colorinlistoftodos, shadow,textsize= footnotesize]{todonotes}
\usepackage{hyperref}

\newcommand{\be}{\begin{equation}}
\newcommand{\ee}{\end{equation}}

\newcommand{\beq}{\begin{eqnarray}}
\newcommand{\eeq}{\end{eqnarray}}

\begin{document}

\title{Parity Symmetry Breaking and Topological Phases in a Superfluid Ring}

\author{Xiurong Zhang}
\affiliation{Institute of Theoretical
Physics and Department of Physics, Shanxi University, 030006,
Taiyuan, China}
\author{Francesco Piazza}
\affiliation{Institut for Theoretische Physik, Universit Innsbruck, A-6020 Innsbruck, Austria}
\author{WeiDong Li}
\thanks{corresponding author: wdli@sxu.edu.cn}
\affiliation{Institute of Theoretical
Physics and Department of Physics, Shanxi University, 030006,
Taiyuan, China}
\author{Augusto Smerzi}
\affiliation{QSTAR, INO-CNR and LENS, Largo Enrico
Fermi 2, I-50125 Firenze, Italy}

\begin{abstract}
We study analytically the superfluid flow of a Bose-Einstein condensate in a ring geometry in presence of a rotating barrier.
We show that a phase transition breaking a parity symmetry among two topological phases occurs at
a critical value of the height of the barrier.
Furthermore,
a discontinuous (accompanied by hysteresis) phase transition is observed
in the ordered phase
when changing the
angular velocity of the barrier.
At the critical point where the hysteresis area vanishes, chemical potential of the ground state develops
a cusp (a discontinuity in the first derivative). Along this path, the jump between the two corresponding states having a different winding number
shows strict analogies with a topological phase transition. We finally study the current-phase relation of the system and compare
some of our calculations with published experimental results.
\end{abstract}


\maketitle

{\it Introduction}.
A paradigmatic manifestation of superfluidity is the existence of stationary atomic states
in a ring geometry in presence of a barrier rotating with constant angular velocity
$\Omega$
\cite{legget_99}. With Bose-Einstein condensates (BEC), these states
have been recently observed
experimentally \cite{Campbell11,moulder12,Boshier13prl,Campbell13,Campbell133}
and extensively studied theoretically
\cite{piazza_2009,mathey_2014,piazza_rings,piazza_crit,rizzi_2014,amico_2015,kavoulakis_2015,yakim_2015,kato_2015,mayol,susanto_2016}.
The stationary current-carrying states are characterized by
a topological invariant
given by the phase of the superfluid accumulated around the ring $\nu=2\pi \ell$, with the integer winding number $\ell= 0, \pm 1,\pm 2...$ \cite{leggett_rmp}.
The winding number can be dynamically modified by sweeping the
angular velocity of the rotating barrier \cite{Campbell13,Campbell133}.
The change in topology takes place via the creation of topological defects (solitons in one dimension $d=1$ \cite{carr_winding} and vortices in $d>1$ \cite{kato_2015,yakim_2015,mayol,piazza_crit} ).

In the limit of a vanishing barrier, the state with topological defects adiabatically connect two rotation-invariant states with different winding number $\ell$. A second-order phase transition takes place two times as a function of $\Omega$ \cite{carr_winding}, first as the system enters the state with topological defects from the first rotational-invariant state $\ell_1$ and then as it leaves the former by entering the second rotational-invariant state $\ell_2$. This scenario changes in presence of any finite-size obstacle that breaks the rotational symmetry of the ring, wherein the topological defects are always dynamically unstable \cite{finazzi_2015}, so that, in general,
two topologically different states cannot be adiabatically connected. This has been recently confirmed experimentally with a barrier moving inside a toroidal BEC \cite{Campbell14Na}, where hysteresis appears in the transition between states with different topological winding number. The unstable branch of the hysteresis loop corresponds to the state with topological defects, and the
angular velocity at which the metastable state decays (through phase slippage \cite{piazza_2009}) into the ground state generalizes
the Landau critical velocity to the weak-link case \cite{finazzi_2015,mayol}.

In this manuscript, we show that with a barrier rotating at the
angular velocity $ \Omega_c= \hbar/2 mR^2$, with $R$ and $m$ the radius of the ring and the atomic mass, respectively,
the ground state of the system becomes degenerate when the height of the barrier is smaller than a critical value $V < V_c$.
The degeneration arises from a parity symmetry breaking that provides two possible ground states with different
topology, i.e., winding number.
In the disordered phase, $V > V_c$, the ground state is unique with an undefined winding number.
Furthermore, by keeping constant the height of the barrier in the ordered phase, $V < V_c$,
a first order phase transition between the two ground states  with different topological winding number
and hysteresis can be observed by varying $\Omega$.
The area enclosed by the hysteresis path
shrinks while increasing the height of the barrier till eventually vanishing at the critical point $V = V_c$.
Hysteresis has been experimentally observed
but the sudden change in the winding number at $\Omega_c$ was smeared out due to shot-to-shot number and finite temperature fluctuations
 \cite{Campbell14Na}.
As order parameter of the phase, both the continuous and discontinuous phase transitions we choose
the
difference between the phase accumulated around the ring $\nu$ and phase drop across the barrier, a quantity which is experimental accessible  \cite{Campbell14prx}. The phase drop across the barrier, together with the current flowing through the
ring, also provides the current-phase relation \cite{Piazza,Campbell14prx} -- an optimal characterization of the ring-superfluid junction \cite{barone,likharev,packard,schwartz70,sols}.

We finally emphasize that at the angular velocity $\Omega_c$ and $V=V_c$,
the transition between the two topological states is accompanied
by a discontinuity in the derivative of the ground state chemical potential
as a function of the angular velocity. Furthermore, at this point the transition is not associated
with the breaking of
any symmetry and it cannot therefore be characterized by
a local order parameter.
This carries strong similarities with a continuous topological
phase transition occurring between two degenerate ground states with different topological winding numbers
$\ell$.

{\it The model}.
We consider a BEC confined in an effective one-dimensional toroidal trap in presence of a
barrier rotating with a constant angular velocity $\Omega$. The barrier is
a penetrable repulsive potential with radial extension larger than the annulus width.
 The system can be modeled by the Gross-Pitaevskii equation (GPE) \cite{pit_str} that governs
the dynamics along the azimuthal coordinate $x\in[-L/2,L/2]$, where $L$ is the length of the ring.
We remove the time-dependence of the Hamiltonian by moving to a rotating reference
frame: $x \Rightarrow x+\Omega R t$
with the torus radius $R=L/2\pi$. This introduces a gauge field $\propto \Omega R$ into the GPE, which reads
\begin{eqnarray} \nonumber
  && i\hbar\frac {\partial}{\partial t} \Psi (x,t)= \left[  \hat{H} +Ng |\Psi (x,t)|^{2} \right] \Psi (x,t), \\
 && \hat{H} = \frac{\hbar^2}{2m}\left(i\frac{\partial}{\partial x}+m \frac{\Omega R}{\hbar}\right)^2+ V(x)-\frac{1}{2}m\Omega^2 R^2 \label{GP}.
\end{eqnarray}
The barrier $V(x)$ is a repulsive square well with height $V>0$ and width $d$ centered about $x=0$,
$N$ is the number of atoms and $g=4\pi\hbar^2a_s/m$ is the contact interaction with the effective 1D s-wave scattering length $a_s$.
With the further transformation $\Psi(x,t)=e^{\imath(m\Omega R x+m\Omega^2 R^2 t/2)/\hbar} \phi(x,t)$,
the gauge field can be removed from the Hamiltonian
which now reads as the usual nonlinear GPE for the order parameter $\phi(x,t)$ \cite{Bransden,Landau}. Following \cite{yu_66,langer_67,schwartz70,carr_05,li06,li04,Piazza}, the stationary solutions of Eq.(\ref{GP}) can be written in terms of Jacobi Elliptical
$\operatorname{SN}$ functions \cite{sup}.
Two class of solutions which we call, for reasons that will become clear below,
plane-waves (PW) and solitons (SL), are found for each value of
the winding number $\ell$. The circulation is $\nu=\oint \Theta(x) dx=2\pi\ell$, where $\Theta(x)=m\Omega R x/\hbar+\theta(x)$ is
the phase in the lab frame while
 $\theta(x)= (m/\hbar) \int dx ~j/\rho(x)$ is the phase in the rotating frame. The BEC density is
 $\rho(x)=|\Psi(x)|^2=|\phi(x)|^2$ and in the rotating frame the current $j$ and the chemical potential $\epsilon$
 are related with the current and chemical potential   in the lab frame
by $I(x)=j+\Omega R\rho(x)$ and $\mathcal{E}=\epsilon-m\Omega^2 R^2/2$, respectively.
In absence of barrier,
$V = 0$,
the current for the PW solution is simply $I=\ell ~ I_0$, where we choose
$I_0=R\Omega_0\rho_0$ and $\Omega_0=\hbar/mR^2$ as units of current and rotation velocity
and a density normalized as $\rho_0=1/L$.
The SL state has a chemical potential larger than the chemical potential of the PW state $\mu_0= N g \rho_0$, which will be used to define
our units of energy, time $\hbar/\mu_0$, and length $\xi_0=\hbar/\sqrt{2m\mu_0}$.
The presence of a repulsive
barrier breaks the rotational invariance and the two solutions at fixed $\Omega,\nu$ are neither purely a PW or a SL. As already mentioned,
we found two kinds of solutions that will
be labeled as PW (SL) since both continuously reduce to an exact PW or a SL as $V\to 0$
\cite{Piazza}.

\begin{figure}[htp]
\includegraphics[width=0.45\textwidth]{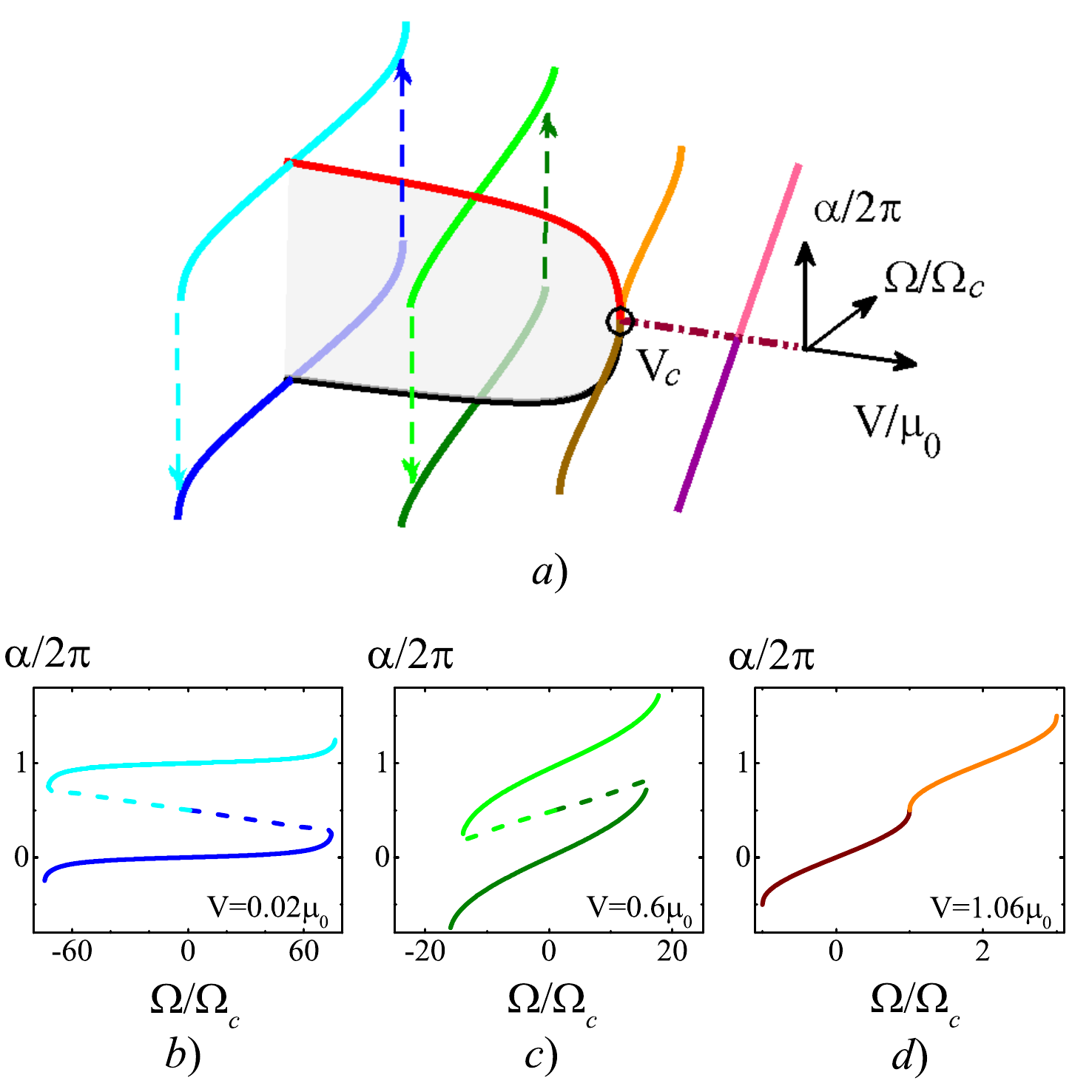}
\caption{a) Order parameter $\alpha$ as a function of the
barrier angular velocity $\Omega$ and height of the barrier $V$.
a) At fixed $\Omega=\Omega_c$, the ground state solution of the system becomes degenerate at $V < V_c$.
The black and red solid line correspond to the value of $\alpha$ for the PW-branch with winding number
$\ell=0$ or $\ell=1$, respectively, plotted as a function of $V$. At $V \ge V_c$, the order parameter vanishes and the winding number of the
state is undefined, dashed-dot line.
Further solid lines running along the $\Omega$ direction for different values of $V$ give $\alpha$ also for the PW-branch, where the different colours correspond to different winding number $\ell=0$ or $\ell=1$.
Hysteresis along the closed trajectories marked by dark-light green and dark-light blue colours exists for $V < V_c$.
b-d) Value of $\alpha$ as a function of $\Omega$ for three different values of $V$. Solid(dashed) lines correspond to the PW (
SL)-branch and different colours correspond to different winding number:
$\ell=0$ or $\ell=1$.
In d), at $V=V_c=1.06\mu_0$, the
$\ell=0$ and the $\ell=1$ PW-branches are directly connected at a point where the derivative of $\alpha$ as a function of $\Omega$ diverges. Here the parameters are the same as in Fig.~\ref{fig1}.}\label{fig2}
\end{figure}

\begin{figure}[htp]
\includegraphics[width=0.45\textwidth]{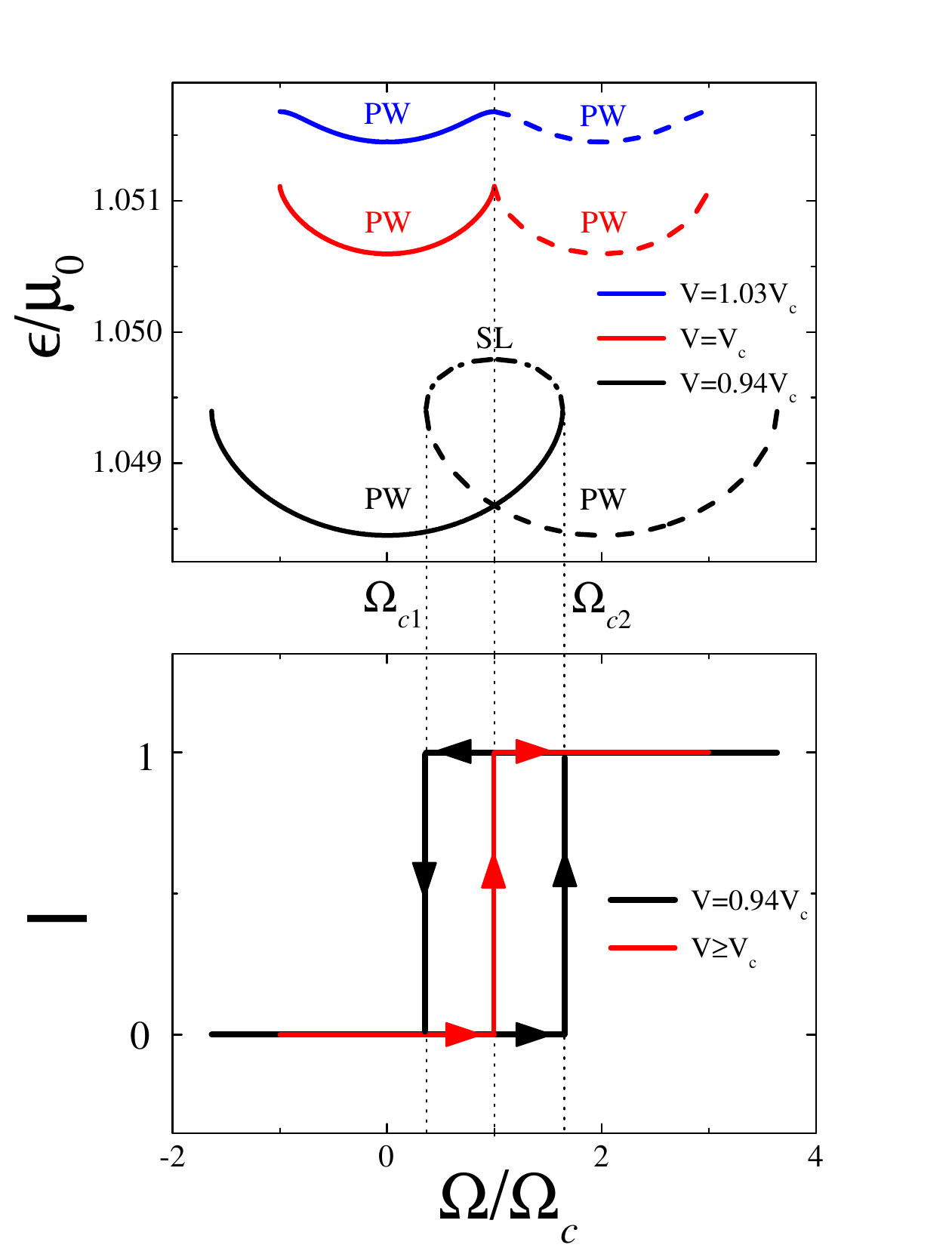}
\caption{Upper panel: chemical potential for three different values of the
barrier height (in units of $\mu_0$). Lower panel: topological winding number below and over the critical
barrier height. Arrows highlight the hysteretic behaviour as a function of the rotation velocity. Here $L=20d$ and $d=20\xi_0$, similarly to the NIST experiment \cite{Campbell14prx}. }\label{fig1}
\end{figure}

{\it Continuous phase transition}.
In the following we study the exact ground state solutions as a function of the order parameter \begin{equation}
\alpha=\nu - \gamma, \label{alf}
\end{equation}
that is the the difference between the circulation $\nu$ and the
the phase drop across the barrier $\gamma$ \cite{sup}. In the limit $V=0$, the phase
difference is simply equal to the
the phase accumulated around the ring: $\alpha= \nu=2 \pi \ell $.
This quantity has been measured experimentally \cite{Campbell14prx} from the interference fringes
of two overlapping BEC, one expanding from a ring with barrier, the second from a disk without barrier
providing the reference phase.

The phase diagram of the system is depicted in Fig.~\ref{fig2}a), where the order parameter $\alpha$ is plotted as a function of the
angular velocity $\Omega$ and strength $V$. When $V < V_c$,  the ground state is 
a PW
with winding number either $\ell=0$ or $\ell=1$
and is characterized by a non-vanishing $\alpha$. This bifurcation is a pitchfork for $\Omega=\Omega_c$, with the unstable branch for $V<V_c$ being the SL solution (not shown in Fig.~\ref{fig2} a), see dashed lines in Fig.~\ref{fig2} b),c)). For $\Omega\neq\Omega_c$ the bifurcation becomes a saddle-node (see the discussion of Fig.~\ref{fig4} below). The behaviour of $\alpha$ as a function of $\Omega$ is shown in Fig.~\ref{fig2} b-d) for three different values of $V$, where the solid (dashed) lines correspond to the PW (SL)-branch and the different colours correspond to different winding numbers.

\begin{figure}[htp]
\includegraphics[width=0.45\textwidth]{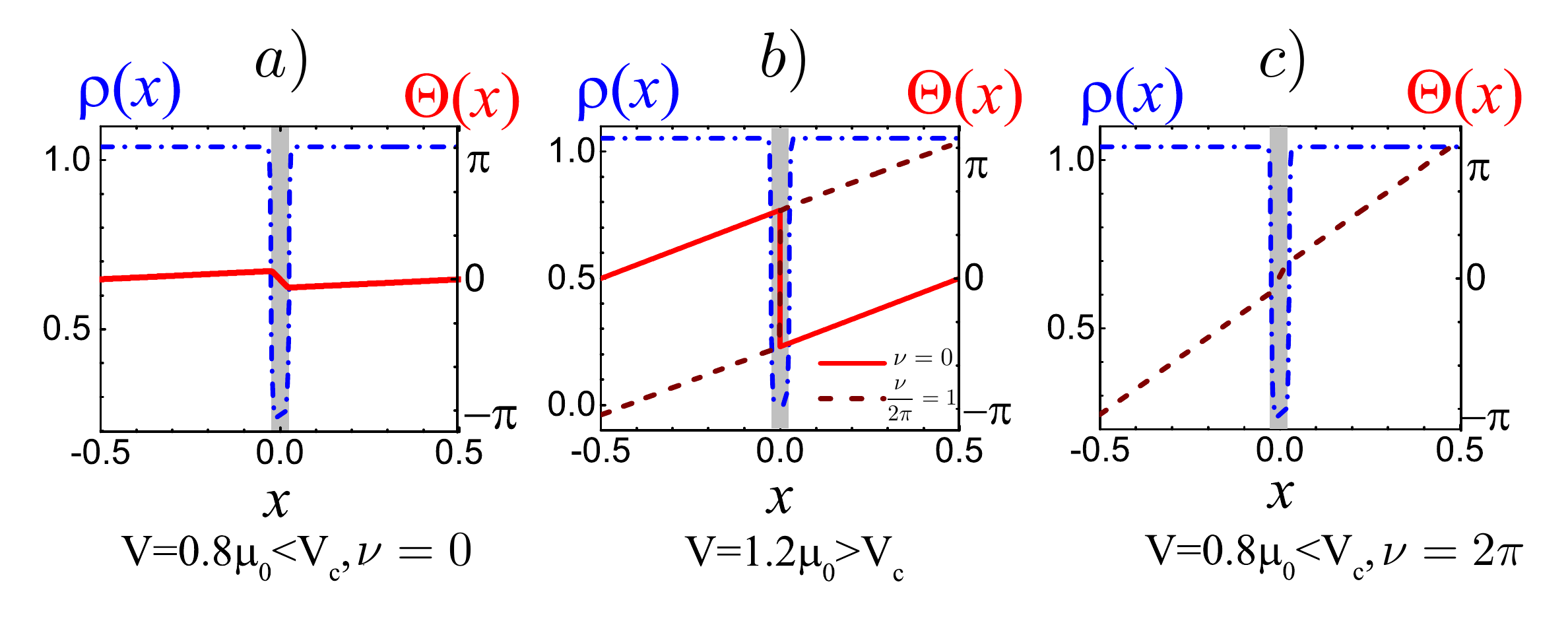}
\caption{Density and phase profiles of the PW-solutions in the hysteretic regime a) and c) and in the non-hysteretic regime b). The shaded area indicates the barrier region. Here the parameters are the same as in Fig.~\ref{fig1}.}\label{fig3}
\end{figure}

It is instructive to analyze how a non-vanishing order parameter $\alpha$ arises by looking at the particular spatial form of the solutions, shown in Fig.~\ref{fig3}. For a fixed angular velocity $\Omega_c$
the behavior of the density and phase of the PW-solution is shown both inside and outside the hysteretic region. In absence of hysteresis: $V\geq V_c$, the $\nu=0$ and $\nu=2\pi$ branches share the same density profile, characterized by a zero at the center of the weak link: $x=0$. At this singular point, the phase has a $\pi$-jump, downards for the $\nu=0$-branch, upwards for the $\nu=2\pi$-branch, leading to the same value of $\alpha$ (see Fig.~\ref{fig2}). For $x\neq 0$ the phase grows linearly with the same slope for both branches. The presence of a singular point (topological defect) in the PW-branches indicates that the latter acquire a solitonic character in the non-hysteretic regime. The SL and PW branches for a given $\nu$ and $\Omega_c$
are indeed equal for $V\geq V_c$ and the winding number
$\ell$ is not defined along this path, dashed dot line in Fig. \ref{fig2} a).

{\it Discontinuos phase transition and hysteresis}.
With a barrier height $V$ below the critical value $V_c$
the system support hysteresis, as already experimentally demonstrated in \cite{Campbell14Na}.
In the region $\Omega<\Omega_{c1}$ the PW state with $\ell=0$ has the lowest energy, while in the region $\Omega>\Omega_{c2}$
the lowest energy state
is a PW with $\ell=1$. In the region $\Omega_{c1}<\Omega<\Omega_{c2}$ either one of the PW solutions is stable while the other is metastable. The metastable PW-branch is connected with the
SL-branch for $\Omega_{c1} \le \Omega \le \Omega_{c2}$, while outside this region only a single PW-branch exists.
The value of $\Omega_{c1,c2}$ are determined by the interaction strength $gN$, the height and the width of the barrier. The fact that the SL-branch in this region is unstable explains the hysteretic behavior \cite{mayol}, see the lower panel of Fig.~\ref{fig1}: as soon as the PW-branch meets the
SL-branch a dynamical instability sets in whereby the system decays into the lowest-energy PW-branch having a different winding number.
This dynamical instability originates from the underlying saddle-node bifurcation where the PW- and the
SL-branch merge \cite{finazzi_2015} (see also Fig.~\ref{fig4}).  We remark that in this case the change of the topological winding number
$\ell$, taking place while going from the metastable to the stable PW-branch, is discontinuous.
The situation changes when $V\geq V_c$: in this case hysteresis is absent and the two PW-branches with $\ell=0$ and $\ell=1$ are directly connected, without the intermediate unstable
SL-branch. Therefore, as shown in Fig.~\ref{fig1}, at $\Omega=\Omega_c$ the topological winding number jumps between
 $\ell=0$ and $\ell=1$, while the system remains in the lowest-energy stationary state. Moreover, as evident from the upper panel of Fig.~\ref{fig1}, if we additionally tune the
 barrier height to $V=V_c$ the chemical potential shows a discontinuous derivative at $\Omega=\Omega_c$. This can be interpreted as
a topological phase transition (a transition between two topologically distinct states) without breaking any local symmetry.
This behaviour is always present, independently
of the particular form of the
barrier.
The disappearance of hysteresis for high enough
barriers has been observed experimentally \cite{Campbell14Na}. Yet the observed transition between states with a different winding number was not perfectly sharp, probably due to shot-to-shot atom-number fluctuations. In order to verify our scenario involving a ``topological"
phase transition one would need to observe both i) a sharp jump between $\ell=0,1$ as a function of $\Omega$ and ii) a second-order discontinuity in some observable (like the chemical potential shown in Fig.~\ref{fig1}). In order to observe i), the temperature has to be low enough to suppress random nucleation of topological defects \cite{mathey_2014} -- as probably already being the case of \cite{Campbell14Na} -- and shot-to-shot number fluctuations need to be reduced. The measurement of a discontinuity in the derivative of the chemical potential as required in ii) seems a more demanding task.

\begin{figure}[htp]
\includegraphics[width=0.45\textwidth]{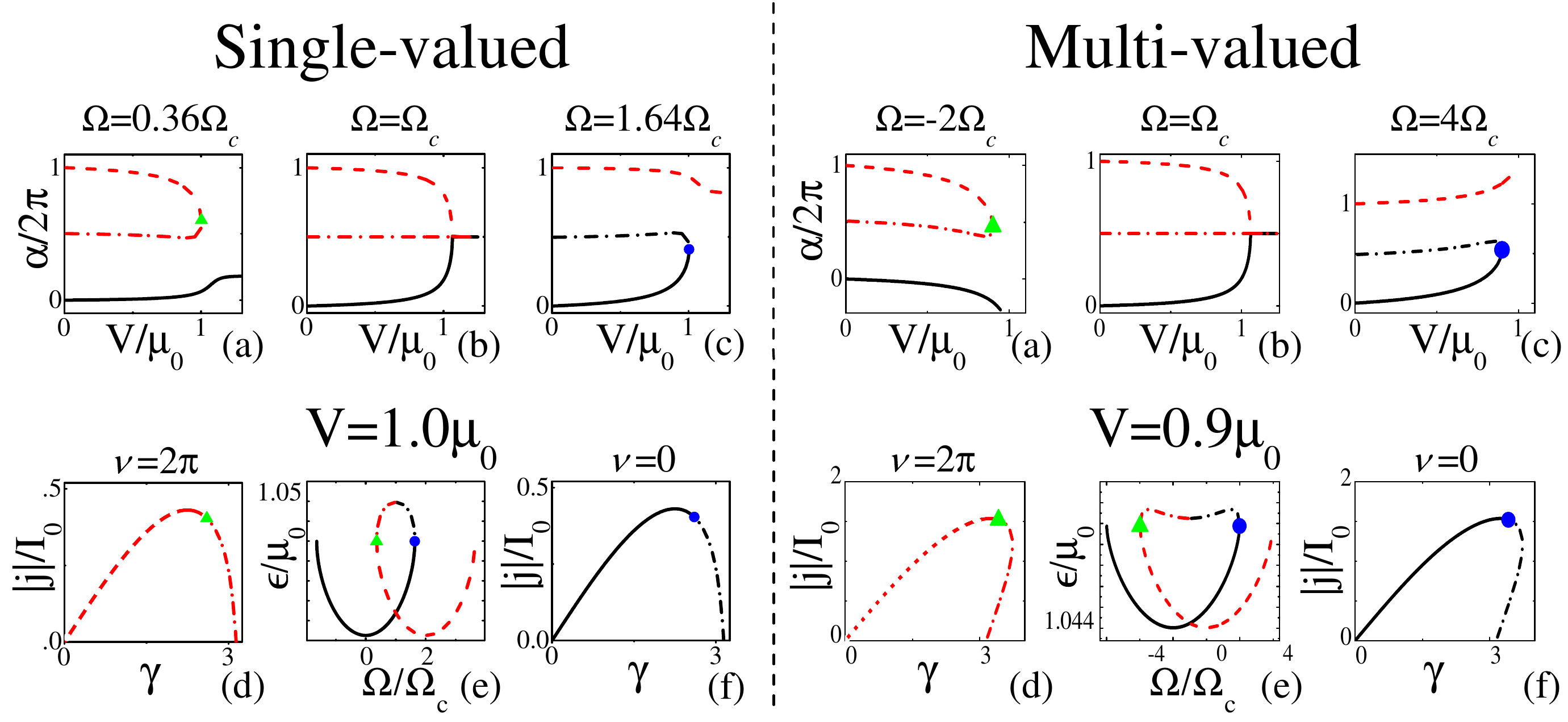}
\caption{Current-phase relation with a rotating barrier. Panels a-c) show the order parameter
 $\alpha$ as a function of the
 barrier height $V$. Panels d),f) show the current-phase relation, while panel e) reports the chemical potential versus
$\Omega$.  In a), d), e) the black solid line represents the $\nu=0$ PW-branch, while the dashed and dash-dotted red line corresponds to the $\nu=2\pi$ PW- and SL branch, respectively. In c),f), the red dashed line represents the $\nu=2\pi$ PW-branch while the solid and dash-dotted black line corresponds to the $\nu=0$ PW- and SL branch, respectively. In b) the
SL branches for $\nu=0,2\pi$ overlap. The blue circle and green triangles mark the special points (saddle-node bifurcations) where the PW- and
SL-branch meet. In the left panel, the current-phase relation is single-valued i.e.
$\gamma<\pi$  while in the right panel is multivalued, namely, for some values of the current $j$ we have
$\gamma>\pi$. Here the parameters are the same as in Fig.~\ref{fig1}.}\label{fig4}
\end{figure}

{\it Current-phase relation}.
The knowledge of the phase drop
$\gamma$ across the barrier, combined with the knowledge of the (spatially-constant) current $j$ flowing across the weak-link,
allows to construct the current-phase relation of the system. This is a powerful characterization of the weak link,
allowing for instance to distinguish different regimes ranging from deep tunneling to hydrodynamic flow \cite{barone,likharev,packard}. In the context of BECs, the current phase-relation has been computed so far for infinite systems with open boundary conditions, a static weak link, and a given injected flow \cite{watanabe,Piazza}.
Stimulated by the experimental results in \cite{Campbell14prx},
we compute here the current-phase relation for our case of a BEC in a ring geometry.
The results are shown in Fig.~\ref{fig4}. For a given
barrier, interaction strength, and winding number, the current-phase relation can be constructed by varying the
angular velocity $\Omega$. As illustrated above, for each fixed $\Omega$, i.e. fixed current $j$, we obtain two solutions (PW and
SL branches) with a different value of
$\gamma$. The current-phase relation for both $\ell=0$ and $\ell=1$ is shown in Fig.~\ref{fig4} d),f) for two different values of the
barrier height $V$. The current-phase relation is composed of the PW- and
SL-branches, meeting at the special points indicated by blue circles or green triangles. The same points are marked also in the $\mu$ versus $\Omega$ diagram (panel e)), as well as in the $\alpha$ versus $V$ diagram (panels a) and c)). It appears how those special points are saddle-node bifurcations, where the PW- and
SL-branch merge and disappear so that there are no stationary solutions for larger (or smaller) values of $V$ or $\Omega$. In b) we also show that at $\Omega=\Omega_c$ the bifurcation becomes a pitchfork, as previously discussed. The latter is characterized by the merging of four branches: the two PW-branches with $\nu=0,2\pi$ (black solid and red dashed lines) and the two
SL-branches with $\nu=0,2\pi$ (red dash-dotted line), which have the same $\alpha$.

\begin{figure}[htp]
\includegraphics[width=0.45\textwidth]{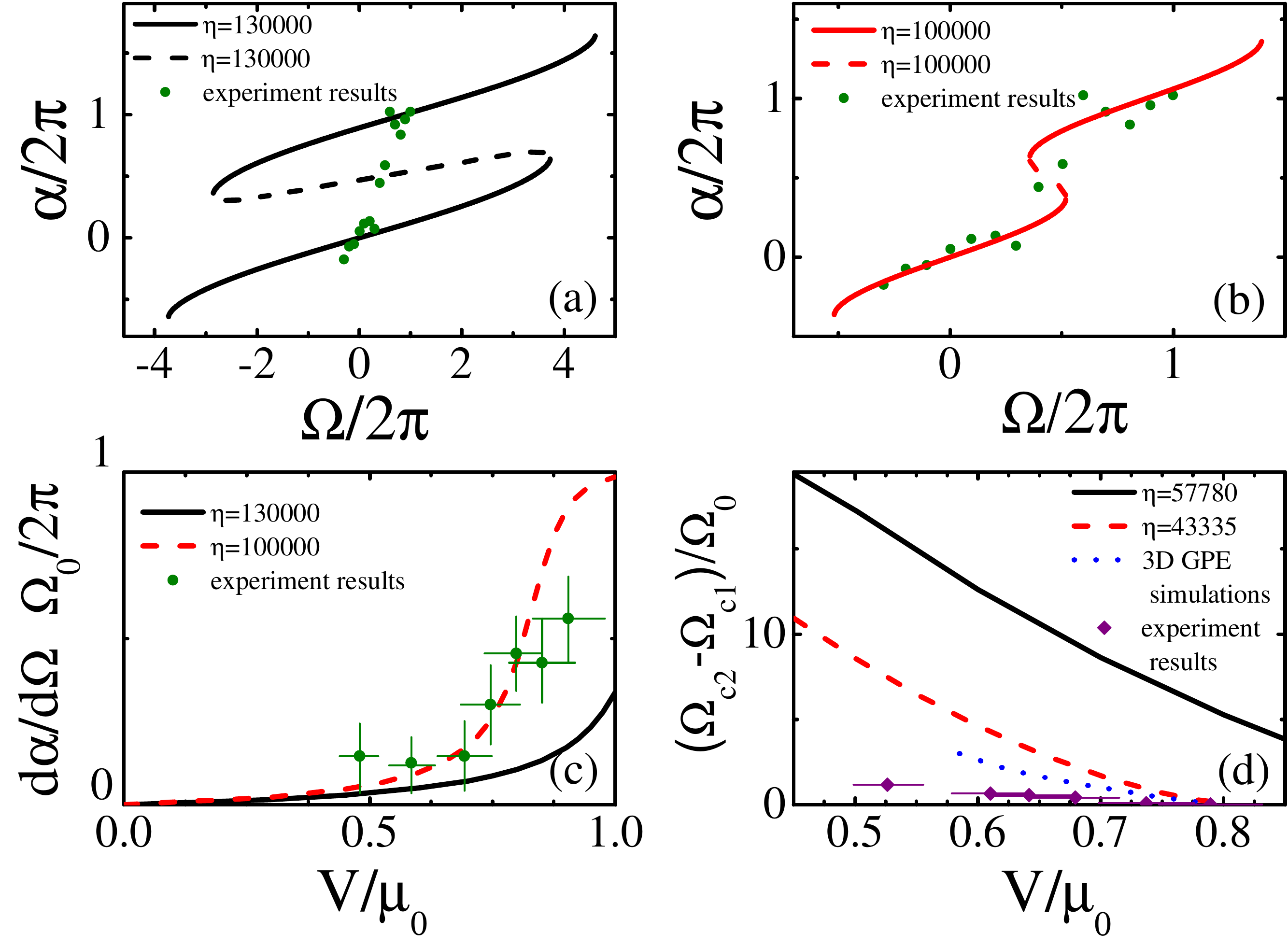}
\caption{Comparison with the experimental measurements of \cite{Campbell14prx} for the order parameter
$\alpha$ as a function of
angular velocity $\Omega$, without a), and with b) the fitted nonlinear parameter $\eta$ (see text).  c), rate of change of $\alpha$ as a function of barrier height $V$. In (d) the size of the hysteresis loop is compared to the value measured in \cite{Campbell14Na} and to the full 3D GPE simulations (blue dotted line) employed in \cite{Campbell14Na}; the black line corresponds to the predictions without fitting parameters, while the red lines to the predictions with a nonlinearity $\eta$ also reduced by $25\%$. In (a) and (b), the
barrier height is $V=0.8\mu_0$. In (a-c), the
barrier width is chosen according to \cite{Campbell14prx} to be $d\approx0.04L\approx22\xi_0$, while in (d) it is taken to be
$d\approx0.05L\approx17\xi_0$, according to \cite{Campbell14Na}.} \label{fig5}
\end{figure}

The current-phase relation indicates the maximal current $j$ and the largest phase drop
$\gamma$ for a given
barrier. In the deep tunneling regime, the current-phase relation is sinusoidal, while in the hydrodynamic regime of flow, achieved for
barriers much smaller than the chemical potential,
the current is quite higher and linearly proportional to the phase drop over a broad range of phases \cite{packard}. Moreover, there is a further regime where the phase drop can be larger than $\pi$, which implies that the current-phase relation becomes multivalued, as shown in the right part of Fig.~\ref{fig4}.

{\it Comparison with experiments}.
All the predictions presented in this manuscript can be experimentally tested within the experimental current state of the art.
In this final section we
compare some of our results with experimental results already obtained at NIST and published in \cite{Campbell14Na,Campbell14prx}.
The comparison is summarized in Fig.~\ref{fig5}. Apart from the
barrier width along the azimuthal coordinate, taken from \cite{Campbell14Na,Campbell14prx}, the most relevant parameter
is the dimensionless effective nonlinearity
\[
\eta=N\times L mg/\hbar^2.
\]
As apparent from Fig.~\ref{fig5}(a) and (b), the agreement between our predictions and the experimental data strongly depends on the value of $\eta$, determined by the total atom number $N$ and ring length $L$. In (a) our predictions are calculated by taking
$N=8\times10^5$ and
$L=140\mu m$ from \cite{Campbell14prx} without any adjustable parameters,
which clearly overestimates the size of the hysteresis loop. However,
 as shown in (b), a very good agreement could be provided, after reducing the effective  nonlinearity $\eta$ by $25\%$, as confirmed in (c) by comparing also the variation of $\alpha$ with the velocity $\Omega$.
 The fact that our purely 1D model overestimates the nonlinearity at given $N,L$ is due to the fact that the experiment is indeed not in the one-dimensional regime. Still we can reproduce the experimental results even quantitatively by simply readjusting the effective nonlinearity. This is consistent with the comparison presented in \cite{Campbell14prx}, where an effective one-dimensional model showed a good agreement once the proper dimensional reduction was performed.

{\it Conclusions}.
We have studied the superfluid flow of a Bose-Einstein condensate confined in a ring geometry in presence of a rotating barrier.
The stationary solutions have been found by solving analytically an effective one-dimensional Gross-Pitaevskii equation.
We have identified a continuous
parity symmetry breaking phase transition among two topological phases.
A discontinuous
phase transition accompanied by hysteresis as a function of the angular velocity of the barrier.
Hysteresis has been experimentally observed at NIST \cite{Campbell14Na,Campbell14prx}.
At the critical point where the hysteresis area vanishes, the chemical potential of  the ground state develops  a cusp (a discontinuity in the first derivative). Along this path, the jump between the two corresponding winding numbers shows strict analogies with a topological phase transition. A good agreement between the order parameter $\alpha$ as a function of the angular velocity and the rate $d\alpha/d\Omega$ as a function
of the height of barrier and the area of the hysteresis has been found with published experimental data in \cite{Campbell14Na,Campbell14prx}
by readjusting the effective nonlinearity to take into account the fact that the experiment is not purely one-dimensional.

{\it Acknowledgments}.
This work was supported by the National Natural Science Foundation of China (Grant No. 11374197), PCSIRT (Grant No. IRT13076) and the Hundred Talent Program of Shanxi Province (2012).

\end{document}